\documentclass[aos,seceqn,nameyear,noautosecdot,dvips]{arximspdf}
\usepackage{dcolumn}
\usepackage{graphicx}

\doi{10.1214/10-AOS814}
\volume{38}
\issue{5}
\pubyear{2010}
\firstpage{2998}
\lastpage{3027}

\makeatletter

\newcolumntype{d}[1]{D{.}{.}{#1}}

\newtheorem{lemma}{Lemma}[section]
\newtheorem{corollary}{Corollary}[section]
\newtheorem{proposition}{Proposition}[section]
\newtheorem{theorem}{Theorem}[section]

\newcommand{\RR}{\mathbb{R}}
\newcommand{\II}{\mathcal I}
\newcommand{\CC}{\mathcal C}
\newcommand{\HH}{\mathcal H}
\newcommand{\KK}{\mathcal K}
\newcommand{\diag}{\operatorname{diag}}

\newcommand{\dom}{\operatorname{dom}}

\makeatother

\begin{document}
\begin{frontmatter}

\title{Quasi-concave density estimation}
\runtitle{Quasi-concave density estimation}

\begin{aug}
\author[A]{\fnms{Roger} \snm{Koenker}\corref{}\thanksref{t1}\ead[label=e1]{rkoenker@uiuc.edu}} and
\author[B]{\fnms{Ivan} \snm{Mizera}\thanksref{t2}\ead[label=e2]{mizera@stat.ualberta.ca}}
\runauthor{R. Koenker and I. Mizera}
\affiliation{University of Illinois and University of Alberta}
\address[A]{Department of Economics\\
University of Illinois\\
410 David Kinley Hall\\
1407 W. Gregory, MC-707\\
Urbana, Illinois 61801\\
USA\\
\printead{e1}}
\address[B]{Department of Mathematical\\
\quad and Statistical Sciences\\
University of Alberta\\
CAB 632, Edmonton, AB\\
T6G 2G1
Canada\\
\printead{e2}}
\end{aug}

\thankstext{t1}{Supported in part by NSF Grant SES-08-50060.}
\thankstext{t2}{Supported by the NSERC of Canada.}

\received{\smonth{8} \syear{2009}}
\revised{\smonth{12} \syear{2009}}

%
\begin{abstract}
Maximum likelihood estimation of a log-concave probability density is
formulated as a convex optimization problem and shown to have an
equivalent dual formulation as a constrained maximum Shannon entropy
problem. Closely related maximum Renyi entropy estimators that impose
weaker concavity restrictions on the fitted density are also
considered, notably a minimum Hellinger discrepancy estimator that
constrains the reciprocal of the square-root of the density to be
concave. A limiting form of these estimators constrains solutions to
the class of quasi-concave densities.
\end{abstract}

%

\end{frontmatter}

\section{\texorpdfstring{Introduction.}{Introduction}}
Our objective is to introduce a general class of shape constraints
applicable to the estimation of probability densities, multivariate as
well as univariate. Elements of the class are represented by
restricting certain monotone functions of the density to lie in convex
cones. Maximum likelihood estimation of log-concave densities
constitutes an important special case; however, the wider class allows
us to include a variety of other shapes. A one parameter subclass
modeled on the means of order $\rho$ studied by \citet{HLP}
incorporates all the quasi-concave densities, that is, all densities
with convex upper contour sets. Estimation methods for these
densities, as described below, bring new opportunities for statistical
data analysis.

Log-concave densities play a crucial role in a wide variety of
probabilistic models: in reliability theory, search models, social
choice and a broad range of other contexts it has proven convenient to
assume densities whose logarithm is concave. Recognition of the
importance of log-concavity was already apparent in the work of
Schoenberg and Karlin on total positivity beginning in the late
1940s. \citet{K68} forged a link between log-concavity and classical
statistical properties such as the monotone likelihood ratio property,
the theory of sufficient statistics and uniformly most powerful tests.
Maximum likelihood estimation of densities constrained to be
log-concave has recently enjoyed a considerable vogue with important
contributions of Walther (\citeyear{wal01}, \citeyear{wal02}, \citeyear{walther08}),
\citet{PalWooMey07}, \citet{Rufi07}, \citet{RD09},
\citet{BRW}, \citet{chawal07} and \citet{CSS}, among others.

Log-concave densities are constrained to exhibit exponential tail
behavior. This restriction motivates a search for weaker forms of the
concavity constraint capable of admitting common densities with
algebraic tails like the $t$ and $F$ families. The $\rho$-concave
densities introduced in Section \ref{s:mle} constitute a rich source of
candidates. While it would be possible, in principle, to consider
maximum likelihood estimation of such densities, duality
considerations lead us to consider a more general class of maximum
entropy criteria. Maximizing Shannon entropy in the dual is equivalent
to maximum likelihood for the leading log-concave case, but other
entropies are also of interest. Section \ref{s:dual} describes several examples
arising in the dual from the class of R\'enyi entropies, each
corresponding to a distinct specification of the concavity constraint,
and each corresponding to a distinct fidelity criterion in the primal.
The crucial advantage of adapting the fidelity criterion to the form
of the concavity constraint is that it assures a convex optimization
problem with a tractable computational strategy.

\section{\texorpdfstring{Quasi-concave probability densities and their
estimation.}{Quasi-concave probability densities and their estimation}}
\label{s:mle}

A probability density function, $f$, is called \textit{log-concave} if
$-{\log f}$ is a (proper) convex function on the support of $f$. We
adhere to the usual conventions of \citet{Roc70}, which allow convex
functions to take infinite values---although we will allow only
$+\infty$, because all our convex functions will be \textit{proper}. The
\textit{domain} of a convex (concave) function, $\dom g$, is then the
set of
$x$ such that $g(x)$ is finite. We adopt the convention $- {\log0} =
+\infty$.

Unimodality of concave functions implies that log-concave densities
are unimodal. An interesting connection in the multivariate case was
pointed out by \citet{Silverman81}: the number of modes of a kernel
density estimate is monotone in the bandwidth when the kernel is
log-concave. However, as illustrated by the Student $t$ family, not
every unimodal density is log-concave. Laplace densities, with their
exponential tail behavior, are; but heavier, algebraic tails are ruled
out. This prohibition motivates a relaxation of the log-concavity
requirement.

\subsection{\texorpdfstring{A hierarchy of $\rho$-concave
functions.}{A hierarchy of $\rho$-concave functions}}

A natural hierarchy of concave functions can be built on the
foundation of the weighted means of order $\rho$ studied
by \citet{HLP}: for any $p$ in the unit simplex,
$\mathcal S= \{ p \in\RR^n | p \geq0, \sum p_i =1 \}$, let
\[
M_{\rho} (a;p) = M_{\rho}(a_1, \ldots, a_n ; p)
= \Biggl(\sum_{i=1}^n p_i a_i^{\rho} \Biggr)^{ 1/\rho},
\]
for $\rho\ne0$; the limiting case for $\rho= 0$ is
\[
M_{0} (a;p) = M_{\rho}(a_1, \ldots, a_n ; p)
= \prod_{i=1}^n a_i^{p_i}.
\]
The familiar arithmetic, geometric and harmonic means correspond to
$\rho$ equal to $1$, $0$ and $-1$, respectively.
Following \citet{A72}, a nonnegative, real function~$f$, defined on
a convex set $C \subset\RR^d$ is called \textit{$\rho$-concave} if for
any $x_0, x_1 \in C$ and $p \in\mathcal S$,
\[
f(p_0 x_0 + p_1 x_1 ) \geq M_\rho(f(x_0), f(x_1); p).
\]
In this terminology, $\log$-concave functions are $0$-concave and
concave functions are $1$-concave. As $M_\rho(a,p)$ is monotone,
increasing in $\rho$ for $a \geq0$ and any $p \in\mathcal S$, it follows
that if $f$ is $\rho$-concave, then $f$ is also $\rho'$-concave for
any $\rho' < \rho$. Thus, concave functions are $\log$-concave, but
not vice-versa. In the limit $-\infty$, concave functions satisfy the
condition
\[
f(p_0 x_0 + p_1 x_1 ) \geq\min\{f(x_0), f(x_1) \},
\]
so they are (and consequently for all $\rho$-concave
functions) \textit{quasi-concave}.

The hierarchy of $\rho$-concave density functions was considered in
the economics literature by \citet{capnal91} in spatial models of
voting and imperfect competition; their results reveal some intriguing
connections to Tukey's half-space depth in multivariate statistics;
see \citet{m02}. Curiously, it appears that the first thorough
investigation of the mathematical concept of quasi-concavity was
carried out by \citet{fin49}. Further details and motivation for
$\rho$-concave densities can be found in \citet{Prekopa},
\citet{Borell} and \citet{DJD}.

\subsection{\texorpdfstring{Maximum likelihood estimation of log-concave
densities.}{Maximum likelihood estimation of log-concave densities}}

Suppose that $X = \{ X_1,\ldots,X_n \}$ is
a collection of data points in $\RR^d$ such that the convex
hull of $X$, $\HH(X)$, has a nonempty interior in $\RR^d$; such a
configuration occurs with probability $1$ if $n \geq d$ and the $X_i$
behave like a random sample from $f_0$, a probability density with
respect to the Lebesgue measure on $\RR^d$. Viewing the $X_i$'s as
a random sample from an unknown, log-concave density $f_0$, we can
find the \textit{maximum likelihood estimate} of $f_0$ by solving
%
\begin{equation}
\label{!lcmle}
\prod_{i=1}^n f(X_i) = \max_f !\qquad
\mbox{such that $f$ is a log-concave density}.
\end{equation}
It is convenient to recast (\ref{!lcmle}) in terms of $g = -{\log f}$,
the estimate becoming $f = {e}^{-g}$,
%
\begin{equation}
\label{!lcmlelog}\qquad
\sum_{i=1}^n g(X_i) = \min_g !\qquad
\mbox{such that $g$ is convex}\quad\mbox{and}\quad \int{e}^{-g(x)}\,
dx = 1.
\end{equation}
The objective function of (\ref{!lcmlelog}) is equal to $+\infty$,
given the convention adopted above, unless all $X_i$ are in the domain
of $g$. As in \citet{sil82}, it proves convenient to move the integral
constraint into the objective function,
%
\begin{equation}
\label{!warden}
\frac{1}{n} \sum_{i=1}^n g(X_i)
+ \int{e}^{-g(x)} \,dx = \min_{g} !\qquad
\mbox{such that $g$ is convex},
\end{equation}
a device that ensures that the
solution integrates to one without enforcing this condition
explicitly. Apart from the multiplier $1/n$, the
crucial difference between (\ref{!lcmlelog}) and (\ref{!warden}) is
that the latter is a convex problem, while the former not.

It is well known that na\"\i ve, unrestricted maximum likelihood
estimation is doomed to fail when applied in the general density
estimation context: once ``log-concave'' is dropped from the
formulation of (\ref{!lcmle}), any sequence of putative maximizers is
attracted to the the linear combination of point masses situated at
the data points. One escape from this ``Dirac catastrophe'' involves
regularization by introducing a roughness, or complexity, penalty;
various proposals in this vein can be found in \citet{Goo71},
\citet{sil82}, \citet{Gu02} and \citet{KoeMiz06}.

Another way to obtain a well-posed problem is by imposing shape
constraints, a line of development dating back to the celebrated
\citet{G56} nonparametric maximum likelihood estimator for monotone
densities. While monotonicity regularizes the maximum likelihood
estimator, unimodality \textit{per se}---somewhat surprisingly---does
not. The desired effect is achieved only by enforcing somewhat more
stringent shape constraints---for instance log-concavity, sometimes
also called ``strong unimodality.'' An advantage of shape constraints
over regularization based on norm penalties is that it is not
encumbered by the need to select additional tuning parameters; on the
other hand, it is limited in scope---applicable only when the shape
constraint is plausible for the unknown density.

\subsection{\texorpdfstring{Quasi-concave density estimation.}{Quasi-concave density estimation}}

Expanding the scope of our investigation, we now replace
${e}^{-g}$ in the
integral of the objective function by a generic function $\psi(g)$ and define
%
\begin{equation}
\label{!probj}
\Phi(g) = \frac{1}{n}\sum_{i=1}^n g(X_i) + \int\psi(g(x))\, dx.
\end{equation}
The following conditions on the form of $\psi$ will be imposed:
\begin{itemize}[(A5)]
\item[(A1)] $\psi$ is a nonincreasing, proper convex function on $\RR$.
\item[(A2)] The domain of $\psi$ is an open interval containing
$(0,+\infty)$.
\item[(A3)] The limit, as $\tau\to+\infty$, of $\psi(y + \tau
x)/\tau$ is
$+\infty$ for every real $y$ and any $x < 0$.
\item[(A4)]
$\psi$ is differentiable on the interior of its domain.
\item[(A5)] $\psi$ is bounded from below by $0$, with
$\psi(x) \to0$ when $x \to+\infty$.
\end{itemize}
The most crucial condition is (A1) ensuring the convexity of $\Phi$.
Condition (A2) assures that $\psi(x)$ is finite for all $x > 0$, while
(A3) is required in the proof of the existence of the estimates. The
relationship between primal and dual formulations of the estimation
problem is facilitated by (A4), and (A5) rules out possible complications
regarding the existence of the integral $\int\psi(g) \,dx$ in
(\ref{!probj}), allowing for the convention $\psi(+\infty)=0$. In the
spirit of the Lebesgue integration theory, the integral then
\textit{exists}, although $\psi(g)$ does not have to be \textit{summable}:
it is either finite [which is automatically true for any $g$ convex
and $\psi(g) = {e}^{-g}$] or $+\infty$. In the latter case, the
objective function $\Phi(g)$ is considered to be equal to $+\infty$;
$\Phi(g)$ is also $+\infty$ if $g(X_i) = +\infty$ for some $X_i$,
which occurs unless all $X_i$ lie in the domain of $g$. On the other
hand, any $g$ equal to some positive constant on $\HH(X)$ and
$+\infty$ elsewhere yields $\Phi(g) < \infty$.

A rigorous treatment without assumption (A5), that is, for functions
$\psi$ not bounded below, would introduce technicalities involving
handling of the integrals in the spirit of singular integrals of
calculus of variations, a strategy resembling the contrivance of
\citet{Huber67} of subtracting a fixed quantity from the objective
function to ensure finiteness of the integral. Although we do not
believe that such formal complications are unsurmountable, we do not
pursue such a development.

Careful deliberation reveals that replacing $g$ by its closure (lower
semicontinuous hull) does not change the integral term in
(\ref{!probj}), and potentially only decreases the first term; this
means that without any restriction of its scope, we may reformulate
the estimation problem as
%
\begin{equation}
\label{!dumb}
\frac{1}{n}\sum_{i=1}^n g(X_i) + \int\psi(g(x)) \,dx =
\min_{g} !\qquad
\mbox{subject to $g \in\KK$},
\end{equation}
where $\KK$ stands for the class of closed (lower semicontinuous)
convex functions on $\RR^d$.

Unlike in (\ref{!warden}), $\psi(g)$ is not necessarily the estimated
density $f$; the relationship of $g$ to $f$ will be revealed in
Section \ref{s:dual}, together with the motivation leading to concrete instances
of some possible functions $\psi$.

\subsection{\texorpdfstring{Characterization of estimates.}{Characterization of estimates}}

We now establish that the estimates, the solutions of (\ref{!dumb}), admit
a finite-dimensional characterization, which is a key to many of their
theoretical properties. For every collection $(X,Y)$ of points
$X_i \in\RR^d$ and $Y_i\in
\RR$, we define a function
%
\begin{equation}
\label{!lch}
g_{(X,Y)}(x) = \inf\Biggl\{ \sum_{i=1}^n \lambda_i Y_i \Bigm|  x =
\sum_{i=1}^n \lambda_i X_i, \sum_{i=1}^n \lambda_i = 1, \lambda_i
\geq0 \Biggr\}.
\end{equation}
Any function of this type is finitely generated in the sense of
\citet{Roc70}, whose Corollary 19.1.2 asserts that it is
polyhedral, being the maximum of finitely many affine functions, and
therefore convex. The convention $\inf\varnothing= +\infty$ used
in (\ref{!lch}) means that the domain of $g_{(X,Y)}$ is equal to $\HH
(X)$. If $h$ is a convex function such that $h(X_i) \leq Y_i$, for all
$i$, then $h(x) \leq g_{(X,Y)}(x)$ for all $x$; the function
$g_{(X,Y)}$ is thus the maximum of convex functions with this
property---the lower convex hull of points $(X_i,Y_i)$.

For fixed $X$, we will denote the collection of all functions
$g_{(X,Y)}$ of the form (\ref{!lch}) by $\mathcal{G}(X)$. The
collection $(X,Y)$ determines $g_{(X,Y)}$ uniquely, by virtue of its
definition (\ref{!lch}). Given $X$, we call a vector $Y$ with
components $Y_i \in\RR$ \textit{discretely convex relative to} $X$,
if there exists a convex function $h$ defined on $\HH(X)$ such that
$h(X_i) = Y_i$. Any function $g$ from $\mathcal{G}(X)$ determines a
unique discretely convex vector $Y_i = g(X_i)$. The converse is also
true: there is a one--one correspondence between $\mathcal{G}(X)$ and
$\mathcal{D}(X) \subseteq\RR^n$, the set of all vectors discretely
convex relative to~$X$.
\begin{theorem}
\label{t:first}
Suppose that assumption \textup{(A1)} holds true. For every convex function $h$
on $\RR^d$, there is a function $g \in\mathcal{G}(X)$ such that
$\Phi(g) \leq\Phi(h)$; the strict inequality holds whenever
$h \notin\mathcal{G}(X)$ and $\HH(X)$ has nonempty interior.
\end{theorem}

The theorem shows that it is sufficient to seek potential solutions of
(\ref{!dumb}) in $\mathcal{G}(X)$; this means, due to the one--one
correspondence of the latter to $\mathcal{D}(X)$, that the
optimization task (\ref{!dumb}) is essentially finite dimensional. The
theorem also justifies the transition to a more convenient
optimization domain in the primal formulation appearing in the next
section.

\section{\texorpdfstring{Duality, entropy and divergences.}{Duality, entropy and divergences}}
\label{s:dual}

The conjugate dual formulation of the primal estimation problem
(\ref{!dumb}) conveys a maximum entropy interpretation and leads us to
several concrete proposals for $\psi$. To conform to existing
mathematical apparatus, we begin by further clarifying the
optimization and constraint functional classes of our primal
formulation. For definitions and general background on convex
analysis, our primary references are Rockafellar (\citeyear{Roc70},
\citeyear{Roc74}) and
\citet{Zei85}; we may also mention \citet{HirLem93} and
\citet{BorLew06}.

\subsection{\texorpdfstring{The primal formulation.}{The primal formulation}}

Hereafter, $\KK(X)$ will denote the cone of closed (lower
semicontinuous) convex functions on $\HH(X)$, the convex hull of $X$.
This cone is a subset of $\CC(X)$, the collection of functions
continuous on $\HH(X)$; it is important that $\CC(X)$ is a linear
topological space, with respect to the topology of uniform
convergence. Note that $\mathcal{G}(X) \subset\KK(X) \subset\CC(X)$.
In view of Theorem \ref{t:first}, any solution of (\ref{!dumb}) is
also the solution of
%
\begin{equation}
\label{!primal}
\frac{1}{n}\sum_{i=1}^n g(X_i) + \int\psi(g(x)) \,dx =
\min_{g\in\mathcal{C}(X)} !\qquad
\mbox{subject to $g \in\KK(X)$},
\end{equation}
and conversely; thus, we will refer to (\ref{!primal}) as our \textit{primal
formulation}.

\subsection{\texorpdfstring{The dual formulation.}{The dual formulation}}

The conjugate of $\psi$ is
\[
\psi^{\ast}(y) = \sup_{x \in\dom\psi} \bigl( y x - \psi(x) \bigr).
\]
Since $\psi$ is nonincreasing, there are no affine functions with
positive slope that minorize the graph of $\psi$, hence
$\psi^{\ast}(y) = +\infty$ for all $y > 0$. If $\psi$ is
differentiable on the (nonempty) interior of its domain, then
$\psi^{\ast}$ can be obtained using differential calculus---as the
\textit{Legendre transformation} of $\psi$; denoting the derivative
$\psi^{\prime}$ by $\chi$, we have
%
\begin{equation}
\label{!legendre}
\psi^{\ast}(y) = y \chi^{-1}(y) - \psi(\chi^{-1}(y)),
\end{equation}
where $\chi^{-1}(y)$ is any solution, $z$, of the equation $\chi(z) =
y$. The (topological) dual of $\mathcal{C}(X)$ is $\mathcal{C}^\ast
(X)$, the space of (signed) Radon measures on $\HH(X)$; its
distinguished element is $P_n$, the empirical measure supported by the
data points $X_i$. The \textit{polar cone} to $\KK(X)$ is
\[
\KK(X)^{-} = \biggl\{ G \in\mathcal{C}^\ast(X)
\Bigm|\int g \,dG \leq0 \mbox{ for all } g \in\KK(X) \biggr\}.
\]
\begin{theorem}
\label{t:dual}
Suppose that assumptions \textup{(A1)} and \textup{(A2)} hold. The strong (Fenchel) dual
of the primal formulation (\ref{!primal}) is
%
\begin{eqnarray}
\label{!dual}
- \int\psi^{\ast}(-f(y)) \,dy = \max_{f} ! \qquad\mbox{subject to }
f = \frac{d(P_n-G)}{dy},\nonumber\\[-8pt]\\[-8pt]
\eqntext{G \in\KK(X)^{-},}
\end{eqnarray}
in the sense that the value, $\Phi(g)$, of the primal objective for
any $g$ satisfying the constraints of (\ref{!primal}), dominates the
value, for any $f$ satisfying the constraints of (\ref{!dual}), of the
objective function in (\ref{!dual}); the minimal value
of (\ref{!primal}) and maximal value of (\ref{!dual}) coincide.
Moreover, there exists $f$ attaining the maximal value
of~(\ref{!dual}). Any dual feasible function $f$, that is, any $f$
satisfying the constraints of (\ref{!dual}) and yielding finite
objective function of (\ref{!dual}), is a probability density with
respect to the Lebesgue measure: $f \geq0$ and $\int f \,dx = 1$. If
condition \textup{(A4)} is also satisfied, then the dual and primal optimal
solutions satisfy the relationship $f = -\psi^{\prime}(g)$.
\end{theorem}

It should be emphasized that the expression of absolute continuity in
(\ref{!dual}) is a requirement on $F = P_n -G$; the dual objective
function is defined as the conjugate to the primal objective function
$\Phi$, and is equal to $-\infty$ for any Radon measure that
is \textit{not} absolutely continuous with respect to the Lebesgue
measure. This is how regularization operates here: only those $F$
qualify for which $P_n$ gets canceled with the discrete component of
$G$. Once $F$ satisfies this requirement, its density integrates to
$1$, as shown in the proof of Theorem \ref{t:dual}. The nonnegativity
for $f$ yielding finite dual objective function is the consequence of
$\psi^{\ast}(-y)$ being infinite for $y < 0$. In practical
implementations, it may be prudent to enforce $f \geq0$ in the dual
explicitly as a feasibility constraint.

\subsection{\texorpdfstring{The interpretation of the dual.}{The interpretation of the dual}}

An immediate consequence of Theorem~\ref{t:dual} is that we can
reformulate the maximum likelihood problem posed in (\ref{!warden}) as
an equivalent maximum (Shannon) entropy problem.
\begin{corollary}
\label{c:mledual}
Maximum likelihood estimation of a log-concave density as posed
in (\ref{!warden}) has an equivalent dual formulation
%
\begin{eqnarray}
\label{!mledual}
- \int f(y) \log f(y) \,dy = \max_{f} ! \qquad\mbox{subject to } f
= \frac{d(P_n-G)}{dy},\nonumber\\[-8pt]\\[-8pt]
\eqntext{G \in\KK(X)^{-},}
\end{eqnarray}
whose solution satisfies the relationship $f = {e}^{-g}$, where
$g$ is the solution of (\ref{!warden}). In particular, the solution
of (\ref{!warden}) satisfies $\int{e}^{-g(x)} \,dx = 1$, therefore
problems (\ref{!lcmlelog}) and (\ref{!warden}) are equivalent.
\end{corollary}

The emergence of the Shannon entropy is hardly surprising---in
view of the well-established connections of maximum likelihood
estimation to the Kullback--Leibler divergence and maximum entropy.
Note that the dual criterion can be also interpreted as choosing the $f$
closest in the Kullback--Leibler divergence to the uniform distribution
on $\HH(X)$, from all $f$ satisfying the dual constraints.

\subsection{\texorpdfstring{R\'enyi entropies.}{Renyi entropies}}

\begin{figure}

\includegraphics{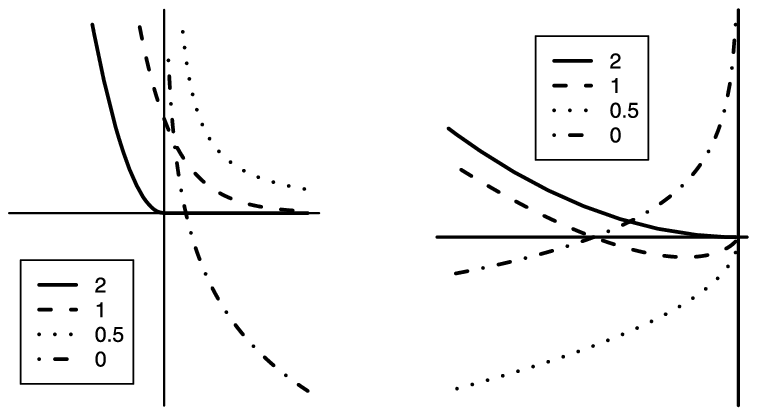}

\caption{Primal $\psi$ (left) and dual $\psi^{\ast}$ (right) for
selected $\alpha\geq0$ from the R\'enyi family of entropies.}
\label{f:1}
\end{figure}

While the outcome of Corollary \ref{c:mledual}, the equivalence
of (\ref{!lcmlelog}) and (\ref{!warden}), could be also shown by
elementary means, it is important to emphasize that the real value of
the dual connection lies in the vista of new possibilities it opens.
To explore the link to potential alternatives, we consider the family
of entropies originally introduced for $\alpha> 0$ by
R\'enyi (\citeyear{Ren61}, \citeyear{Ren65}),
%
\begin{equation}
\label{!rent}
(1 - \alpha)^{-1} \log\biggl( \int f^\alpha(x) \,dx \biggr),\qquad
\alpha\ne1,
\end{equation}
as an extension of the limiting case for $\alpha= 1$, the Shannon
entropy. For $\alpha\ne1$, maximizing (\ref{!rent}) over $f$ is
equivalent to the maximization of
%
\begin{equation}
\label{!rsign}
\frac{\mbox{sgn} (1 - \alpha)}{\alpha} \int f^\alpha(x) \,dx
= - \mbox{sgn} (\alpha-1)\int\frac{f^\alpha(x)}{\alpha} \,dx .
\end{equation}
The dependence of convexity/concavity properties of $y^\alpha$
necessitates a separate treatment of the cases with $\alpha> 1$, when
the conjugate pair is
\[
\psi(x) = \cases{
(-x)^{\beta}/\beta, &\quad for $x \leq0$, \cr
0, &\quad for $x > 0$,}\qquad
\psi^{\ast} (y) = \cases{
(-y)^{\alpha}/\alpha, &\quad for $y \leq0$, \cr
+\infty, &\quad for $y > 0$,}
\]
and the cases with $\alpha< 1$, where
\[
\psi(x) = \cases{
+\infty, &\quad for $x \leq0$, \cr
-x^{\beta}/\beta, &\quad for $x > 0$,}\qquad
\psi^{\ast} (y) = \cases{
-(-y)^{\alpha}/\alpha, &\quad for $y \leq0$, \cr
+\infty, &\quad for $y > 0$,}
\]
where $\beta$ and $\alpha$ are conjugates in the usual sense that
$1/\beta+ 1/\alpha= 1$. See Figure \ref{f:1}.

The general form of the primal formulations (\ref{!primal})
corresponding to
(\ref{!rent}) can be written, for $\alpha\ne1$, in a unified way as
%
\begin{equation}
\label{!genprim}
\frac{1}{n}\sum_{i=1}^n g(X_i) + \frac{1}{|\beta|}
\int|g(x)|^{\beta} \,dx = \min_{g\in\mathcal{C}(X)} !
\end{equation}
together with the relation between the dual and primal solutions, $f =
|g|^{\beta-1}$. Several particular instances merit special attention.

\subsection{\texorpdfstring{Power divergences.}{Power divergences}}

For $\alpha> 1$, we may write $(-g)$ instead of $|g|$, and then
introduce $h = -g$. The resulting primal formulation is
%
\begin{equation}
\label{!renpr}
\qquad - \frac{1}{n}\sum_{i=1}^n h(X_i) + \frac{1}{\beta}
\int h^{\beta}(x) \,dx = \min_{h \in\mathcal{C}(X)} !\qquad
\mbox{subject to $h \in\KK(X)$}.
\end{equation}
By Theorem \ref{t:first}, this formulation is equivalent to
%
\begin{equation}
\label{!renprgen}
- \frac{1}{n}\sum_{i=1}^n h(X_i) + \frac{1}{\beta}
\int h^{\beta}(x) \,dx = \min_{h} !\qquad
\mbox{subject to $h \in\KK$}.
\end{equation}
After substituting $f^{1/(\beta-1)}$ for $h$, multiplying by $\beta$,
and rewriting in terms of $\alpha$ we obtain a new objective function
%
\begin{equation}
\label{!basu}
- \biggl(\frac{\alpha}{\alpha-1}\biggr)
\frac{1}{n} \sum_{i=1}^n f^{\alpha-1}(X_i) + \int f^{\alpha}(x) \,dx,
\end{equation}
which recalls the ``minimum density power divergence estimators,''
proposed, for $\alpha\geq1$, by \citet{BasHarHjo98} in the context
of estimation in parametric families.

\subsection{\texorpdfstring{Pearson $\chi^2$.}{Pearson $\chi^2$}}\label{sec36}
Although $\alpha= 2$ is a special case of the power divergence family
mentioned above, it deserves a special mention. The choice of
$\alpha=2$ in the R\'enyi family leads to the dual formulation
%
\begin{equation}
\label{!rendual}
- \int f^2 (y) \,dy = \max_{f} !\qquad \mbox{subject to }
f = \frac{d(P_n-G)}{dy},\qquad G \in\KK(X)^{-}.\hspace*{-35pt}
\end{equation}
The primal formulation can be written, after the application of
Theorem \ref{t:first}, in a particularly simple form
%
\begin{equation}
\label{!renprim}
\frac{1}{n}\sum_{i=1}^n g(X_i) + \frac{1}{2} \int g^2 (x) \,dx =
\min_{g} ! \qquad\mbox{subject to $g \in\KK$},
\end{equation}
which can be interpreted as a variant of the minimum Pearson $\chi^2$
criterion. A similar theme can be found in the dual, which can
be interpreted as returning among all densities satisfying its
constraints the one with minimal Pearson $\chi^2$ distance to the
uniform density on $\CC(X)$.

The relation between primal and dual optimal solutions is $f = -g$;
the convexity constraint on $g$ therefore implies that $f$ must be
concave. Replacing $g$ in (\ref{!renprim}) by $-f$ and appropriately
modifying the cone constraint gives a variant of the ``least-squares
estimator,'' studied by \citet{GJW01b} and going back at least to
\citet{BirMas93}; the estimate was defined to estimate a \textit{convex}
(and decreasing) density on $\RR^+$, a domain that is apparently still
under the scope of Theorem \ref{t:first}.

\subsection{\texorpdfstring{Hellinger.}{Hellinger}}
While the form of the objective function for $\alpha= 2$ has some
computational advantages, its secondary consequence---constraining the
density itself to be concave rather than its logarithm---is not at all
appealing. Indeed, all R\'enyi choices with $\alpha> 1$ impose a more
restrictive form of concavity than log-concavity. From our
perspective, it seems more reasonable to focus attention on weaker
forms of concavity, corresponding to $\alpha\leq1$. Apart from the
celebrated log-concave case $\alpha= 1$, a promising alternative
would seem to be R\'enyi entropy with $\alpha= 1/2$. This choice in
the R\'enyi system leads to the dual
%
\begin{equation}
\label{!dhell}
\int\sqrt{f(y)} \,dy = \max_{f} ! \qquad\mbox{subject to }
f = \frac{d(P_n-G)}{dy},\qquad G \in\KK(X)^{-},\hspace*{-32pt}
\end{equation}
and primal, again after the application of Theorem \ref{t:first},
%
\begin{equation}
\label{!phell}
\frac{1}{n}\sum_{i=1}^n g(X_i) +
\int\frac{ 1 }{\raisebox{0.4ex}{$g(x)$}} \,dx =
\min_{g} ! \qquad\mbox{subject to $g \in\KK$}.
\end{equation}
The estimated density satisfies $f = 1/g^{2}$, which means that the
primal constraint, $g \in\KK$, enforces the convexity of $g =
1/\sqrt{f}$. In the terminology of Section \ref{s:mle}, the estimated density is
now required to be only $-1/2$-concave, a significant relaxation of the
log-concavity constraint; in addition to all log-concave densities,
all the Student $t_\nu$ densities with $\nu\geq1$ satisfy this
requirement. The dual problem (\ref{!dhell}) can be interpreted as a
Hellinger fidelity criterion, selecting from the cone of dual feasible
densities the one closest in Hellinger distance to the uniform
distribution on $\HH(X)$.

\subsection{\texorpdfstring{The frontier and beyond\textup{?}}{The frontier and beyond}}

Although the original R\'enyi system was confined to $\alpha> 0$, a
limiting form for $\alpha= 0$ can be obtained similarly to the
$\alpha= 1$ case. It yields the conjugate pair
\begin{eqnarray*}
\psi(x) &=& \cases{
+\infty, &\quad for $x \leq0$, \cr
-1/2 - \log x, &\quad for $x > 0$,}
\\
\psi^{\ast} (y) &=& \cases{
-1/2 - \log(-y), &\quad for $y < 0$, \cr
+\infty, & \quad for $y \geq0$.}
\end{eqnarray*}
As is apparent from Figure \ref{f:1}, this $\psi$ violates our condition (A5),
but may nevertheless deserve a brief consideration. Note first that
the possible complications with existence of integrals may occur only
in the formulation (\ref{!dumb}) with unbounded domain---not
in (\ref{!primal}), where all integrals are of bounded functions over
a compact domain. The major technical complications with $\psi$
violating (A5) concern theorems in Section \ref{s:exfish}, and are briefly
discussed there. Here we mention only that the resulting dual, adapted
directly from (\ref{!dual}), is
\[
\int\log{f(y)} \,dy = \max_{f} ! \qquad\mbox{subject to }
f = \frac{d(P_n-G)}{dy},\qquad G \in\KK(X)^{-},
\]
and the primal becomes
\[
\frac{1}{n}\sum_{i=1}^n g(X_i) - \int\log g(x) \,dx =
\min_{g\in\mathcal{C}(X)} !\qquad
\mbox{subject to $g \in\KK(X)$}.
\]
In this case $g = 1/f$, and the estimate is constrained to be
$-1$-concave, a yet still weaker requirement that admits all of the
Student $t_\nu$ densities for $\nu> 0$.

If we interpret the dual problem (\ref{!mledual}), for $\alpha= 1$,
as choosing a constrained $f$ to minimize the Kullback--Leibler
divergence of $f$ from the uniform distribution on $\HH(X)$, we can
similarly interpret the $\alpha= 0$ dual as minimizing the reversed
Kullback--Leibler divergence. In parametric estimation, the latter
objective is sometimes associated with empirical likelihood, while the
former is associated with exponentially tilted empirical likelihood.
See, for example, \citet{HP99} for related discussion in the context
of kernel density estimation, and \citet{Schennach07}.

One might try to continue in this fashion marching inexorably toward
weaker and weaker concavity requirements. There appears to be no
obstacle in considering $\alpha< 0$; the general
form (\ref{!genprim}) of the primal is still applicable. The shape
constraints corresponding to negative $\alpha$ encompass a wider and
wider class of quasi-concave densities, eventually arriving at the
$-\infty$-concave constraint, at which point we would have sanctioned
all of the quasi-concave densities. But formal complications, as well
as computational difficulties dictate the more prudent strategy of
restricting attention to $\alpha> 0$ cases.

\section{\texorpdfstring{Existence and Fisher consistency of
estimates.}{Existence and Fisher consistency of estimates}}
\label{s:exfish}

Returning to our general setting, existence, uniqueness and Fisher consistency
are established under mild conditions on the function $\psi$.

\subsection{\texorpdfstring{Existence of estimates.}{Existence of estimates}}

Theorem \ref{t:first} not only justifies the choice of the
optimization domain in (\ref{!primal}), but also shows, due to the
one--one correspondence between $\mathcal{G}(X)$ and $\mathcal{D}(X)$,
that the optimization task (\ref{!primal}) is essentially
finite dimensional, parametrized by the values $Y_i=g(X_i)$. This
facilitates the proof of the following existence result.
\begin{theorem}
\label{t:ex}
Suppose that assumptions \textup{(A1), (A2), (A3)} and \textup{(A5)} hold, and
that $\HH(X)$ has a nonempty interior. Then the formulation
(\ref{!dumb}) has a solution $g \in\mathcal{C}(X)$; if $\psi$ is
strictly convex, then this solution is unique.
\end{theorem}

\subsection{\texorpdfstring{Fisher consistency.}{Fisher consistency}}

In our general setting, a comprehensive asymptotic theory for the
proposed estimators remains a formidable objective. Considerable
recent progress has been made on theory for the univariate log-concave
($\alpha= 1$) maximum likelihood estimator: \citet{PalWooMey07}
proved consistency in the Hellinger metric, \citet{RD09} prove
consistency in the supremum norm on compact intervals, and
\citet{BRW} derive asymptotic distributions. For maximum likelihood
estimators in $\RR^d$, \citet{CS10} establish consistency for
estimators of a log-concave density, and \citet{SerWel09} for
estimators of convex-transformed densities. These results are surely
suggestive of the plausibility of analogous results for other $\alpha$
and dimensions greater than one. However, the highly technical nature
of the proofs, and their strong reliance on special features of the
univariate setting indicate that such a development may not be
immediate.

While anything else in this direction may be viewed as speculative,
Fisher consistency, a crucial prerequisite for a more detailed
asymptotic theory, can be verified in a quite straightforward manner
and essentially complete generality. For differentiable $\psi$,
Theorem \ref{t:dual} gives the relationship between the solution $g$
of the optimization task (\ref{!primal}) and the density estimate: $f
= -\psi^{\prime}(g)$. Using the notation $\chi$ for $\psi^{\prime}$,
and $\chi^{-1}$ for its inverse, as in Section \ref{s:dual}, we may write $g
= \chi^{-1}(-f)$, and subsequently rewrite the
formulation (\ref{!dumb}) in terms of the estimated density $f$
(omitting, for brevity, the integration variables)
%
\begin{eqnarray}
\label{!primf}
\int\chi^{-1}(-f)\, dP_n + \int\psi(\chi^{-1}(-f)) \,dx =
\min_{f} !\nonumber\\[-8pt]\\[-8pt]
\eqntext{\mbox{subject to $\chi^{-1}(-f) \in\KK$}.}
\end{eqnarray}
This yields a new objective function---which we nevertheless denote,
slightly abusing the notation, also $\Phi$. The population version of
this $\Phi$ is obtained by replacing $ dP_n$ by $f_0 \,dx$:
%
\begin{equation}
\label{!popsi}
\Phi_0(f) =
\int\chi^{-1}(-f) f_0 + \psi(\chi^{-1}(-f)) \,dx.
\end{equation}
The Fisher consistency for an estimator defined by solving
(\ref{!primf}) requires that $\Phi_0(f_0) \leq\Phi_0(f)$, for every
$f$; however, there may be a formal problem now with the existence of
the integral in (\ref{!popsi}), as $\chi^{-1}(f)$ may take both
positive or negative values. A possible way of handling this obstacle
is the strategy of \citet{Huber67}, briefly mentioned in Section \ref{s:mle}:
instead of $\Phi$, we consider a modified objective function
%
\begin{equation}
\label{!primhub}
\tilde{\Phi}(f) =
\int\biggl(\chi^{-1}(-f) + \frac{\psi^{\ast}(-f_0)}{f_0} \biggr) \,dP_n
+ \int\psi(\chi^{-1}(-f)) \,dx,
\end{equation}
which, when minimized over $f$ satisfying the constraint
of (\ref{!primf}), yields an optimization problem equivalent
with (\ref{!primf}), since the difference of $\Phi$ and $\tilde{\Phi
}$ is
constant in $f$. However, the population version of $\tilde{\Phi}$
%
\begin{equation}
\label{!pophub}
\tilde{\Phi}_0(f) =
\int\chi^{-1}(-f) f_0 + \psi^{\ast}(-f_0) + \psi(\chi^{-1}(-f)) \,dx
\end{equation}
is now better suited for the ensuing version of the Fisher consistency
theorem.
\begin{theorem}
\label{t:fish}
Suppose that $\psi$ satisfies assumptions \textup{(A1), (A2), (A4)} and \textup{(A5)}. The
integrand in (\ref{!pophub}) is then nonnegative for any probability
density $f$ such that $\chi^{-1} (-f) \in\KK$, and identically equal
to $0$
for $f=f_0$; therefore,
$0 = \tilde{\Phi}_0(f_0) \leq\tilde{\Phi}_0(f)$,
where $\tilde{\Phi}_0(f)$ is well defined for every $f$, possibly
equal to $+\infty$.
\end{theorem}

In fact, Theorem \ref{t:fish} can be proved in the same manner for the
unmodified $\Phi$, if $\dom\psi= (\omega, +\infty)$ with $\omega>
-\infty$. Then the inverse of $\chi= \psi^{\prime}$, and hence the
range of $\chi^{-1}$ is bounded from below by $\omega$. In such a
case, $\chi(f)f_0 \geq\omega f_0$, so the first term
in (\ref{!popsi}) is minorized by an integrable function $\omega f_0$;
the second term is bounded from below by $0$ by assumption (A5), so the
whole integral then exists in the Lebesgue sense, being either finite
or equal to $+\infty$.

If, however, assumption (A5) is not satisfied, then the existence of the
integral should be assumed explicitly; we return to this point briefly
at the end of the proof of Theorem \ref{t:fish}. Note that, by
comparing (\ref{!legendre}) and (\ref{!popsi}), existence of the
integral is equivalent to assuming the integrability (summability) of
%
\begin{equation}
\label{!legenf0}
f_0 \chi^{-1}(f_0) + \psi(\chi^{-1}(-f_0))
= - \psi^{\ast}(-f_0),
\end{equation}
that is, the existence and finiteness of the entropy term in the dual
(\ref{!dual}).

\section{\texorpdfstring{Examples of practical use.}{Examples of practical use}}
\label{s:data}

We employed two independent algorithms for solving the convex
programming problems posed above: \texttt{mskscopt} from the MOSEK
software package of \citet{And06}, and the \texttt{PDCO} MATLAB
procedure of \citet{saunders04}. Both algorithms are coded in MATLAB
and employ similar primal-dual, log-barrier methods. Further details
regarding numerical implementation appear in Appendix \ref{appB}. The crux of
both algorithms is a sequence of Newton-type steps that involve
solving large, very sparse least squares problems, a~task that is very
efficiently carried out by modern variants of Cholesky decomposition.
Several other approaches have been explored for computing
quasi-concave density estimators that are log-concave. An active set
algorithm for univariate log-concave density estimation was described
in \citet{DumHusRuf07} and implemented in the R package
\texttt{logcondens} of \citet{logcondens}. \citet{CSS} have recently
implemented a promising steepest descent algorithm for multivariate
log-concave estimation that may be adaptable to other quasi-concave
density estimation problems.

\subsection{\texorpdfstring{Univariate example: Velocities of bright
stars.}{Univariate example: Velocities of bright stars}}

\begin{figure}[b]

\includegraphics{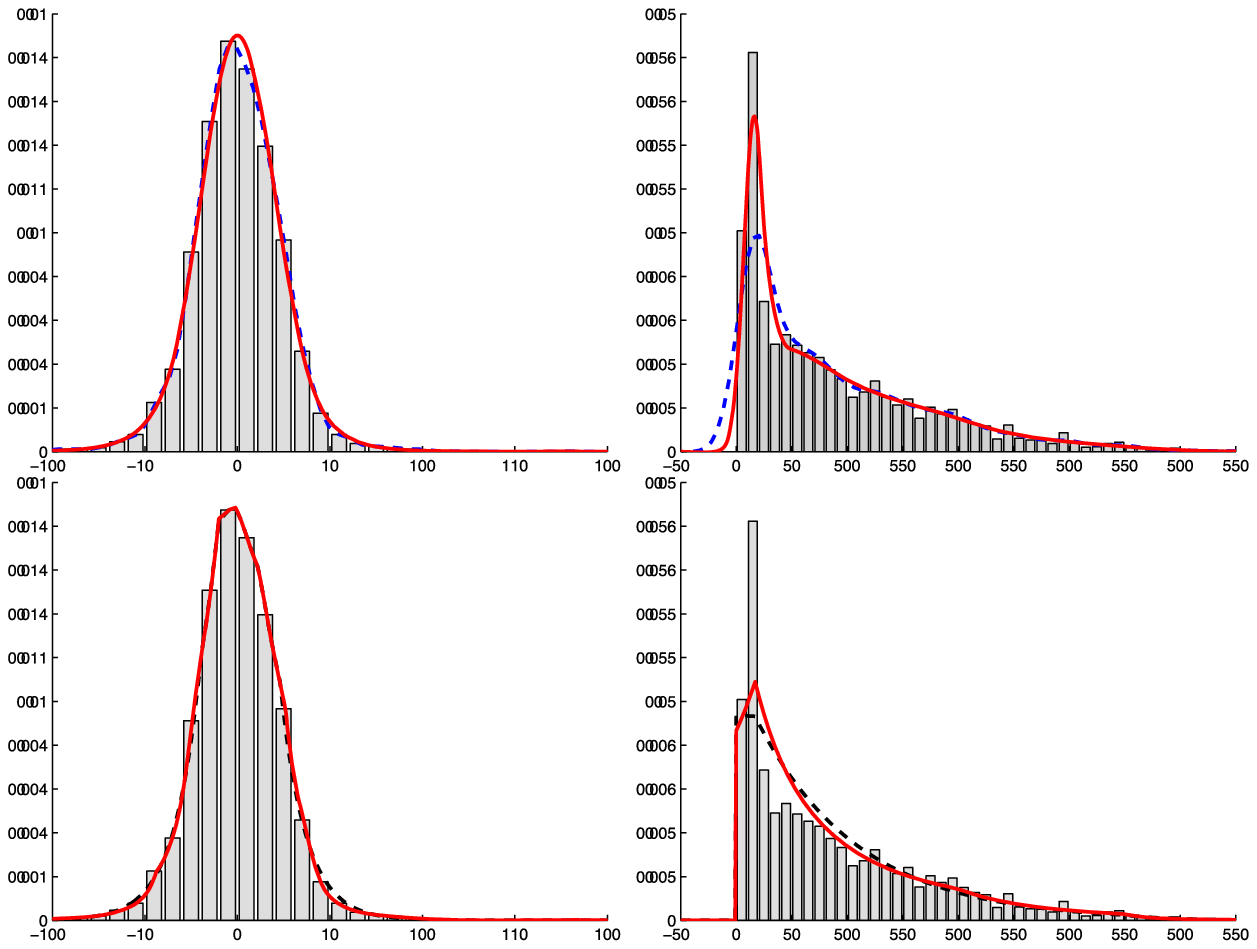}

\caption{The estimates of the densities of radial (left) and
rotational (right) velocities of the stars from the Bright Star
Catalog. Broken lines are kernel density estimates in the upper two
panels, and the solid lines are total variation penalized estimates.
In the lower two panels the broken lines are the log-concave
estimates and the solid lines represent the Hellinger
($-1/2$-concave) estimates.}
\label{figstars}
\end{figure}

To illustrate the application of the foregoing methods, we briefly
consider some realistic examples. Our first example features data
similar to those considered by \citet{PalWooMey07}, the type of
data where shape constraints sometimes arise in a natural manner. The
two samples consist of 9092 measurements of radial and 3933 of
rotational velocity for the stars from Bright Star Catalog,
\citet{hofwar91}. The left and right panels of
Figure \ref{figstars} show the results for the radial and rotational
velocity samples, respectively.

The broken line in the upper panels shows kernel density estimates,
each time with default MATLAB bandwith selection; the solid lines
correspond to one of the norm penalized estimates proposed in
\citet{KoeMiz06}: maximum likelihood penalized by the total variation
of the second derivative of the logarithm of the estimated density.
This is the $L^1$ version of the Silverman's (\citeyear{sil82}) estimator
penalizing the squared $L^2$ norm of the third derivative. The
smoothing parameter for the latter estimate was set quite arbitrarily
at $1$; it seems that this arbitrary choice works quite satisfactorily
here, providing---somewhat surprisingly, for both samples---about the
same level of smoothing as the kernel estimator. For the radial
velocity sample, the two estimates are essentially the same. For the
rotational velocity sample, however, the right upper panel shows that
the kernel density estimate differs substantially from the penalized
one. Both estimators have the unfortunate effect of assigning
considerable mass to negative values despite the fact that there are
no negative observations. This effect is somewhat more pronounced for
the kernel estimate.

Since the preliminary analyses of the upper panels indicates that the
hypothesis of unimodality is plausible for both of the datasets, a
natural next step is the application of a shape-constrained
estimator---a move that, among other things, may relieve us of
insecurities related to the arbitrary choice of smoothing parameters.
The broken line in the lower panels of Figure \ref{figstars} shows the
log-concave maximum likelihood ($\alpha=1$), and the solid line the
Hellinger ($-1/2$-concave) estimate ($\alpha=1/2$). While, as expected,
there is almost no difference between the two (and in fact, among all
four) estimates for the radial velocity dataset, the right lower panel
reveals that the log-concave estimate yields for the rotational
velocity sample a density that is monotonically decreasing---which
contradicts the evidence suggested by all other methods. The Hellinger
estimate, on the other hand, exhibits a subtle, but visible bump at
the location of the plausible mode, thus turning out to be visually
more informative about the center of the data than the tails. This is
somewhat paradoxical given its original heavy-tail motivation,
confirming that the real universe of data analysis can be much more
subtle than that of the surrounding theoretical constructs.

\subsection{\texorpdfstring{Bivariate example: Criminal fingers.}{Bivariate example: Criminal fingers}}

To illustrate our approach in a simple bivariate setting, we
reconsidered the well-known \citet{MacD02} data on the heights and
left middle finger lengths of 3000 British criminals. This data was
employed by Gosset in preliminary simulation work described
in \citet{student}.

%
\begin{figure}

\includegraphics{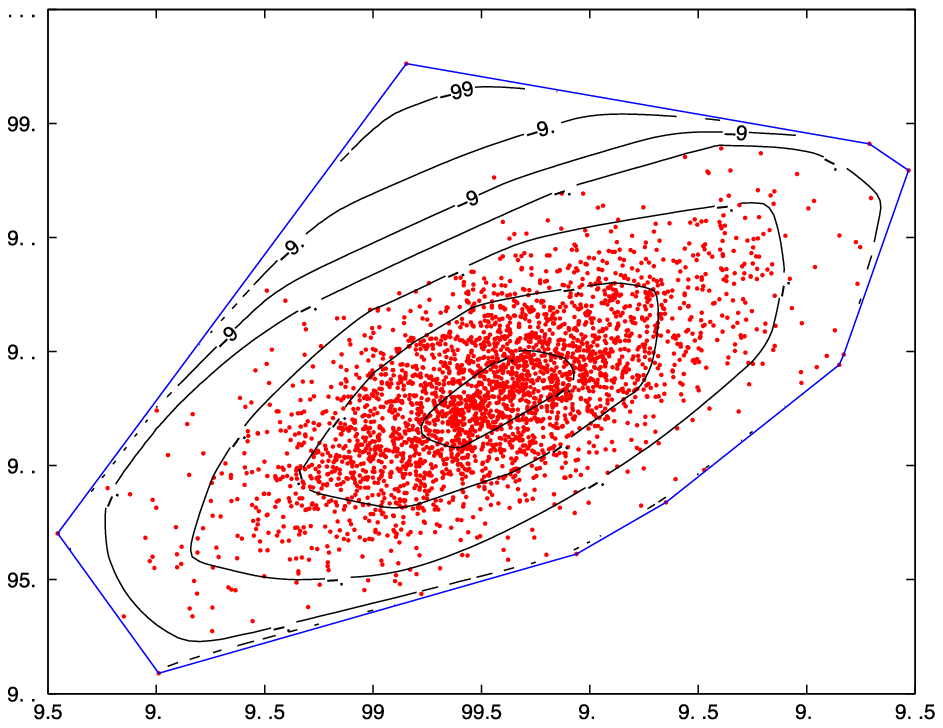}

\caption{Hellinger ($-1/2$-concave) estimate of the density of Student's
criminals. Contours are labeled in units of log-density.}
\label{fig7}
\end{figure}

%
\begin{figure}[b]

\includegraphics{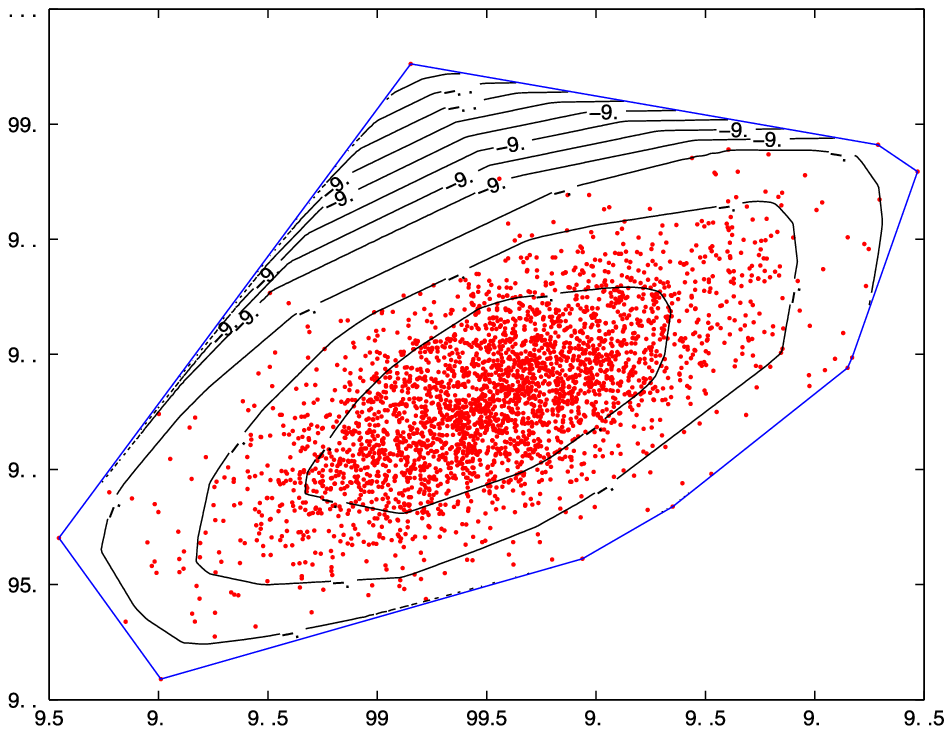}

\caption{Log-concave estimate of the density of Student's criminals.
Contours are labeled in units of log-density.}
\label{fig8}
\end{figure}

Figure \ref{fig7} illustrates the Hellinger ($-1/2$-concave) fit of this
data. Contours are labeled in units of log-density. A notable feature
of the data is the unusual observation in the middle of the upper
edge. This point is highly anomalous, at least for any density with
exponential tail behavior. The maximum likelihood estimate of the
log-concave model in Figure \ref{fig8} has very similar central
contours, but the outer contours fall off much more rapidly implying
that the log-concave estimate assigns much smaller probability to the
region near the unusual point.

\subsection{\texorpdfstring{Some simulation evidence.}{Some simulation evidence}}

Motivated by a suggestion of one of the referees, we undertook some
numerical experiments to explore performance of our shape constrained
estimators and evaluate whether consistency appeared to be a plausible
conjecture. For the log-concave estimator Pal, Woodroofe and Meyer
(\citeyear{PalWooMey07}) report ``Hellinger error'' for a fully crossed design involving
five target densities and five sample sizes with 500 replications per
cell.

\begin{figure}

\includegraphics{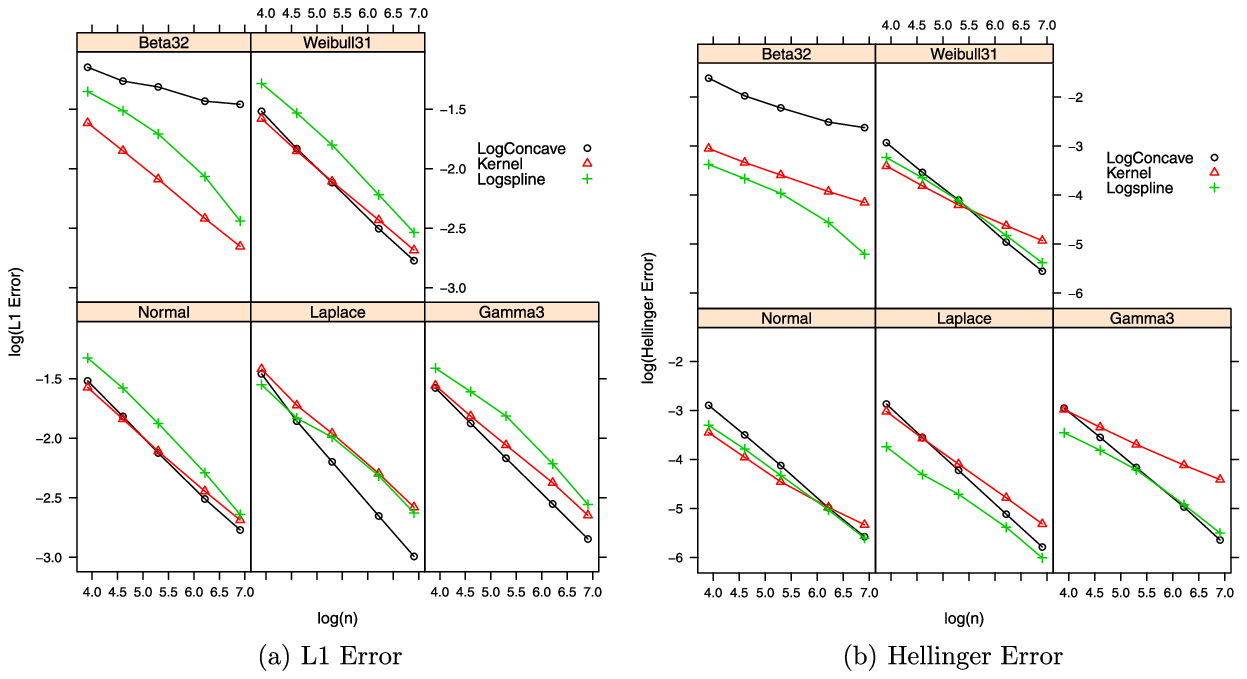}

\caption{Comparison of estimators of several log-concave densities.}
\label{logc}
\end{figure}

In Figure \ref{logc}, we report results of our attempt to reproduce the
PWM experiment expanded somewhat to consider two competing estimators:
the adaptive kernel estimator of Silverman (\citeyear{S86}) using a Gaussian
kernel, and the logspline estimator of Kooperberg and Stone (\citeyear{KS91}) as
implemented in the \texttt{logspline} R package of Kooperberg. Five
target densities are considered: (standard) normal, Laplace, Gamma(3),
Beta($3,2$) and Weibull($3,1$) as in PWM. Five sample sizes are studied:
50, 100, 200, 500, 1000. And two measures of performance are
considered: squared Hellinger distance as in PWM in the left panel and
$L_1$ distance in the right panel. Plotted points in these figures
represent cell means. Both figures support the contention that the
rates of convergence are comparable for all three estimators.

\begin{figure}

\includegraphics{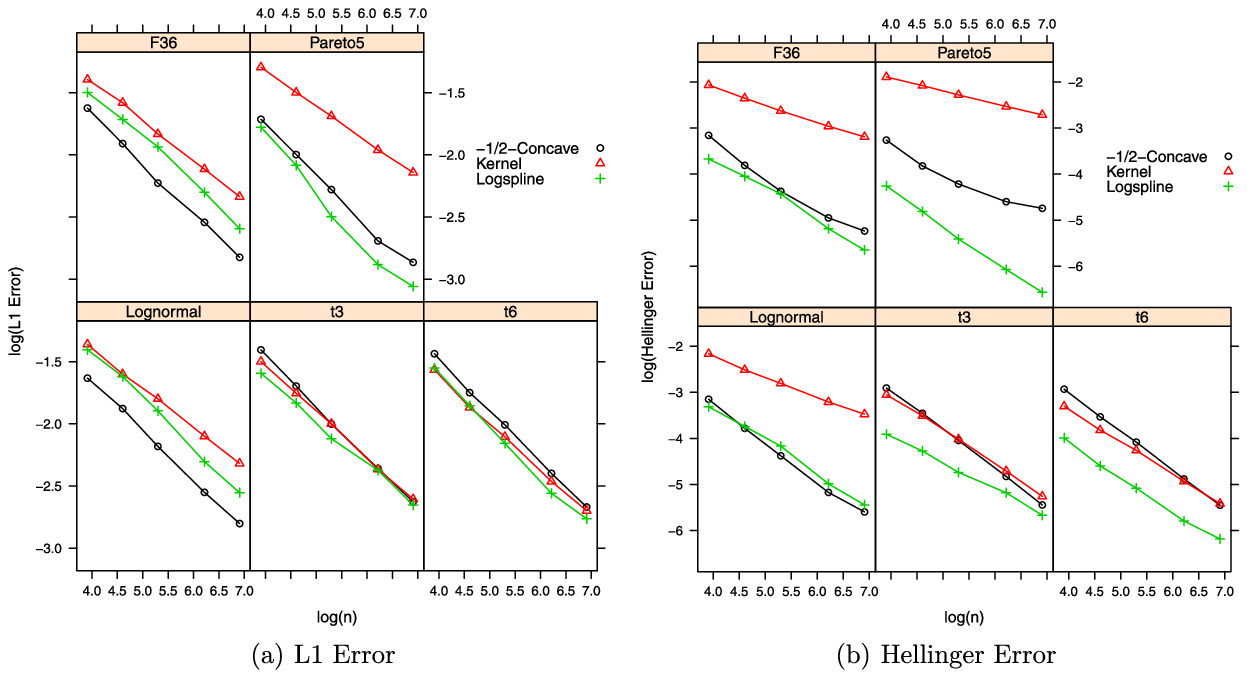}

\caption{Comparison of estimators of several $-1/2$-concave densities.}
\label{hellc}
\end{figure}

Figure \ref{hellc} reports results from a similar experiment for the
the $-1/2$-concave estimator described in Section \ref{sec36}. We consider
five new target densities: lognormal, $t_3$, $t_6$, $F_{3,6}$ and
Pareto(5), all of which fall into the $-1/2$-concave class. The same
competing estimators and sample sizes are used. In a small fraction of
cases for the second group of densities, less than 0.2 percent, there
were problems either with the convergence of the logspline or
shape-constrained estimator, or with the numerical integration
required to evaluate the performance measures, so Figure \ref{hellc}
plots cell medians rather than cell means. Again, the figures support
the conjecture that the rates of convergence for the shape-constrained
estimator are competitive with those of the adaptive kernel and
logspline estimators.

\begin{table}[b]
\caption{Estimated convergence rates for log-concave target
densities}
\label{PWMTab}
\begin{tabular*}{\tablewidth}{@{\extracolsep{\fill}}ld{2.4}d{2.4}d{2.4}@{}}
\hline
\multicolumn{1}{@{}l}{\textbf{Criterion}} & \multicolumn{1}{c}{\textbf{Log-concave}}
& \multicolumn{1}{c}{\textbf{Kernel}} & \multicolumn{1}{c@{}}{\textbf{Logspline}}\\
\hline
L1 error& -0.417 & -0.366 & -0.393\\
& (0.018) & (0.003) & (0.012)\\
Hellinger& -0.875 & -0.498 & -0.698\\
& (0.032) & (0.031) & (0.021)\\
\hline
\end{tabular*}
\end{table}

A concise way to summarize results from these experiments is to
estimate the simple linear model
\[
\log(y_{ij}) = \alpha_i + \beta\log(n_j) + u_{ij},
\]
where $y_{ij}$ denotes a cell average of our two error criteria for
one of our three estimators, for target density $i$ and sample size
$n_j$. In this rather na\"{\i}ve framework, $\hat\beta$ can be interpreted
as an empirical rate of convergence for the estimator. In Tables~\ref{PWMTab} and
\ref{HellTab}, we report these estimates suppressing the estimated target density
%
\begin{table}
\caption{Estimated convergence rates for $-1/2$-concave target
densities}\label{HellTab}
\begin{tabular*}{\tablewidth}{@{\extracolsep{\fill}}ld{2.4}d{2.4}d{2.4}@{}}
\hline
\multicolumn{1}{@{}l}{\textbf{Criterion}} & \multicolumn
{1}{c}{$\bolds{-1/2}$\textbf{-concave}} & \multicolumn{1}{c}{\textbf{Kernel}}
& \multicolumn{1}{c@{}}{\textbf{Logspline}}\\
\hline
L1 error & -0.405 & -0.324 & -0.386\\
& (0.004) & (0.008) & (0.01)\\
Hellinger & -0.751 & -0.355
& -0.672\\
&(0.034) &(0.023) &(0.019)\\
\hline
\end{tabular*}
\end{table}
specific $\alpha_i$'s. In this comparison too, the shape constrained
estimators perform quite well.

\section{\texorpdfstring{Extensions and conclusions.}{Extensions and conclusions}}
\label{s:con}

We have described a rather general approach to qualitatively
constrained density
estimation. Log-concave densities are an important target class, but
other, weaker,
concavity requirements that permit algebraic tail behavior are also of
considerable practical interest. Ultimately, the approach accommodates
all quasi-concave densities as a limit of the R\'enyi entropy family.

There are many unexplored directions for future research. As we have seen,
a consequence of the variational formulation of our
concavity constraints is that the estimated densities vanish off the convex
hull of the data. Various treatments for this malady may be suggested.
\citet{MullerRufi07} have recently suggested applying one of several
estimators of the Pareto tail index to the smoothed ordinates from the
log-concave preliminary density estimator. Our inclination would be to prefer
solutions that impose further regularization on the initial problem. Thus,
for example, we can add a new penalty term to the primal problem, penalizing
the total variation of the derivative (gradient) of $\log f$,
and choosing a suitable value of the
associated Lagrange multiplier to smooth the tail
behavior at the boundary.

We have adhered, thus far, to the principle that the entropy choice
in the fidelity criterion of the dual problem should dictate the form
of the convexity constraint: likelihood thus implies log-concavity,
Hellinger fidelity implies $1/\sqrt{f}$ concavity, etc.
One may wish to break this linkage and consider
maximum likelihood estimation of general $\rho$-concave densities.
This may have some advantages from an inference viewpoint, at the
cost of complicating the numerical implementation.

Embedding shape constrained density estimation of the type considered
here into semiparametric methods would seem to be an attractive option
in many settings. And, it would obviously be useful to consider
inference for the validity of shape constraints in the larger context
of penalized density estimation. We hope to pursue some of these
issues in future work.

\begin{appendix}
\section{Proofs}

\vspace*{-8pt}

\begin{pf*}{Proof of Theorem \ref{t:first}}
Given $h$ convex, put $Y_i = h(X_i)$ and take $g = g_{(X, Y)}$, the
function defined by (\ref{!lch}). The convexity of $h$ implies that
$h(x) \leq g(x)$ for every $x$; since $\psi$ is nonincreasing, we have
%
\begin{equation}
\label{!intineq}
c = \int\psi(h(x)) \,dx - \int\psi(g(x)) \,dx \geq0.
\end{equation}
The definition of $g_{(X,Y)}$ implies that $h(X_i) = g_{(X,Y)}(X_i) =
Y_i$; therefore, the rest of $\Phi$ remains unchanged,
and (\ref{!intineq}) implies that $\Phi(g) \leq\Phi(h)$.

Suppose that $h \notin\mathcal{G}(X)$. Then $h \ne g$ and the
inequality $h \leq g$ implies that $\dom g \subseteq\dom h$. If $\dom
h \ne\dom g$, then there is a point $x_0 \notin\dom g$ from the
interior of $\dom h$, because $\dom{g}$ is closed. The continuity of
$h$ at $x_0$ implies that $h(x) \leq K < +\infty= g(x)$ for all $x$
from an open neighborhood of $x_0$; this proves that $c > 0$. If $\dom
h = \dom g = \HH(X)$, then the polyhedral character of $\HH(X)$
implies through Theorem 10.2 of \citet{Roc70} that $h$ is upper
semicontinuous relative to $\HH(X)$ at any $x \in\HH(X)$; that is, if
$h(x_0) < g(x_0)$ for some $x_0 \in\dom h$, then the inequality holds
for all $x$ in an open, relatively to $\HH(X)$, neighborhood of $x_0$.
Such a relative neighborhood has positive Lebesgue measure, due to the
fact that the interior of $\HH(X)$ is nonempty. Hence, we have $c > 0$
also in this case and the strict inequality $\Phi(g) < \Phi(h)$
follows.
\end{pf*}
\begin{pf*}{Proof of Theorem \ref{t:dual}}
We use the conventional notation $\langle\ell, x \rangle$ to denote
$\ell(x)$, the result of the application of a linear functional to
$x$. The definition of the conjugate of a convex function $H$ in this
notation is
\[
H^{\ast}(y) = \sup_{x \in\dom F} \bigl( \langle y,x \rangle- F(x) \bigr);
\]
the resulting function is convex itself, being a sup of affine
functions. For any $f \in\CC(X)$ and any Radon measure $G$, a linear
functional from $\CC(X)^{\ast}$, we have
\[
\langle G, f \rangle= \int f dG.
\]
We start the proof by rewriting (\ref{!primal}) as
\[
\frac1 n \sum_{i=1}^n g(X_i) + \int\psi(g(x)) \,dx
+ \iota_{\mathcal{K}(X)}(g) = \Phi(g) + \Upsilon(g) = \inf_g !,
\]
where $\Phi$ is the original objective function of (\ref{!primal}) and
$\Upsilon= \iota_{\mathcal{K}(X)}$ is the indicator function of
$\mathcal{K}(X)$. The expression for the Fenchel dual of this type of
problem follows from \citet{Roc66}; see also \citet{Roc74},
Section 5, Example~11:
\[
-\Phi^{\ast}(G) - \Upsilon^{\ast}(-G) = \max_{G} !
\]
(note that one of the conjugates, in both cites sources, is in the
``concave'' sense, which explains the negative sign of the argument in
the second term, but not in the first). The conjugate of the indicator
of a convex cone $\mathcal{K}(X)$ is the indicator of
$-\mathcal{K}(X)^{-}$ [\citet{Roc74}, Section 3, equation (3.14)]. The term
$-\Upsilon^{\ast}(-G)$ in the objective can be therefore interpreted
as a constraint $-G \in-\mathcal{K}(X)^{-}$, that is, $G \in
\mathcal{K}(X)^{-}$. The definition of the conjugate of $\Phi$ gives
%
\begin{eqnarray}
\label{!one}
\Phi^{\ast}(G) &=&
\sup_g \Biggl( \langle G, g \rangle
- \frac1 n \sum_{i=1}^n g(X_i) - \int\psi(g(x)) \,dx \Biggr)
\nonumber\\[-8pt]\\[-8pt]
&=& \sup_g \biggl( \langle G-P_n, g \rangle- \int\psi(g(x)) \,dx \biggr)
= \Psi^{\ast}(G-P_n);\nonumber
\end{eqnarray}
the sup is taken over all $g$ from
\[
\dom\Phi
= \biggl\{ g \in\mathcal{C}(X) \Bigm|\int\psi(g(x)) \,dx < +\infty\biggr\}
= \dom\Psi,
\]
where $\Psi$ is the functional given by
%
\begin{equation}
\Psi(g) = \int\psi(g(x)) \,dx,
\end{equation}
and $\Psi^{\ast}$ is its conjugate. The form of the latter is given by
\citet{Roc71}, Corollary 4A: if $G$ is absolutely continuous with
respect to the Lebesgue measure, then
%
\begin{equation}
\label{!two}
\Psi^{\ast}(G) = \int\psi^{\ast}\biggl(\frac{dG}{dx}\biggr) \,dx,
\end{equation}
otherwise $\Psi^{\ast}(G) = +\infty$. These facts, and expressions
(\ref{!one}) and (\ref{!two}), yield (\ref{!dual}).

Rockafellar [(\citeyear{Roc66}), Theorem 1; see also Rockafellar
(\citeyear{Roc74}), Section 8, Example 11$'$]
gives also a constraint qualification for this type of problem: to
prove strong duality, we need to find some $g$ where both $\Phi$ and
$\Upsilon$ are finite and one of them is continuous. Such a $g$ is
provided by a function constant on $\HH(X)$, say $g(x) = 1$ for all
$x \in\HH(X)$. It is convex, thus $\Upsilon(g) = 0$ is finite. So is
$\Phi(g)$; the topology on $\mathcal{C}(X)$ is that of uniform
convergence, and $\psi$ is continuous at $1$, hence there is a
neighborhood of $g$ containing only functions for which $\Phi$ is
finite and $\Phi$ is continuous at $g$.

Once the constraint qualification is verified, we know that the primal
and dual optimal values coincide (zero duality gap), and that the dual
is attained---there is an optimal solution to the dual; see Theorem
52.B(3) of \citet{Zei85}. Due to the fact that $\psi$ is decreasing,
$\psi^{\ast}(-f) = +\infty$ whenever $f < 0$; thus, if $f$ yields a
finite dual objective function, then $f$ is nonnegative. If
$G \in\KK(X)^{-}$, then $\langle G, f \rangle\leq0$ for every
$f \in\KK(X)$; consequently,
\[
0 \geq\langle G, 1 \rangle= - \langle G, -1 \rangle\geq0.
\]
Therefore, $\langle G, 1 \rangle= 0$ and for every dual feasible $f$,
\[
\int f(x) \,dx = \langle P_n - G, 1 \rangle=
\langle P_n, 1 \rangle- \langle G, 1 \rangle= \int1 \,dP_n = 1.
\]
That is, every dual feasible $f$ is a probability density with respect
to the Lebesgue measure.

If a primal solution, $g$, exists---the fact that is established by
Theorem \ref{t:ex}, but here we are exploring only the consequences of
such a premise---it is related to the dual solution, $f$, via
extremality (Karush--Kuhn--Tucker) conditions. The form of this
relationship asserted by the theorem follows from the second condition
of (8.24) in Rockafellar [(\citeyear{Roc74}), Section 8, Example 11$'$], together with
the form
of the subgradient of $\Psi$ given by Rockafellar [(\citeyear{Roc71}), Corollary 4B],
combined with the fact established above that the estimated density
$f$ corresponds to $F = P_n - G$.
\end{pf*}
\begin{pf*}{Proof of Theorem \ref{t:ex}}
By Theorem \ref{t:first}, any potential solution of (\ref{!dumb}) lies
within the class $\mathcal{G}(X)$ of polyhedral functions; due to the
one--one correspondence between $\mathcal{G}(X)$ and $\mathcal{D}(X)$,
the set of vectors discretely convex relative to $X$, the existence
proof can be carried for (\ref{!primal}) reparametrized by $Y_i$, the
putative values of $g(X_i)$. In what follows, $X$ remains fixed, and
$\alpha$, $\beta$ will denote generic coefficients of convex
combinations: any real numbers satisfying $\alpha, \beta\geq0$,
$\alpha+ \beta= 1$.

As the correspondence between $\mathcal{G}(X)$ and $\mathcal{D}(X)$ is
not a linear mapping (except for $d=1$), the first thing to be shown
is that (\ref{!primal}) remains a convex problem when reparametrized
in terms of a vector $Y \in\RR^n$, with components $Y_1, \ldots, Y_n$.
The resulting problem minimizes, over $Y \in\RR^n$, the objective
function
\begin{eqnarray*}
\Phi_{\mathcal{D}}(Y) &=&
\frac{1}{n}\sum_{i=1}^n Y_i + \int\psi\bigl(g_{(X,Y)}(x)\bigr)
\,dx\qquad \mbox{if $Y \in\mathcal{D}(X)$}
\\
&=& +\infty\qquad\mbox{otherwise}.
\end{eqnarray*}
Note first that $\mathcal{D}(X)$ is a convex subset of $\RR^n$: if $Y,
Z \in\mathcal{D}(X)$, then there exist convex functions $g$, $h$
satisfying $Y_i = g(X_i)$ and $Z_i = h(X_i)$; subsequently, $\alpha
Y_i + \beta Z_i = \alpha g(X_i) + \beta h (X_i)$, and as $\alpha g +
\beta h$ is also a convex function, we obtain that $\alpha Y + \beta Z
\in\mathcal{D}(X)$ for any convex combination of $Y$ and $Z$. Thus,
it is sufficient to show that $\Phi_{\mathcal{D}}$ is convex on
$\mathcal{D}$, which amounts to demonstrating the convexity of the
function $Y \mapsto\int\psi(g_{(X, Y)}(x)) \,dx$. Let $Y, Z \in
\mathcal{D}(X)$; the definition (\ref{!lch}) gives for their convex
combination
\begin{eqnarray*}
g_{(X,\alpha Y + \beta Z)}(x) &=&
\inf\Biggl\{ \sum_{i=1}^n \lambda_i (\alpha Y_i + \beta Z_i)
\Bigm| x = \sum_{i=1}^n \lambda_i X_i,
\sum_{i=1}^n \lambda_i = 1, \lambda_i \geq0 \Biggr\}
\\
&\geq&\alpha\inf\Biggl\{ \sum_{i=1}^n \lambda_i Y_i
\Bigm| x = \sum_{i=1}^n \lambda_i X_i,
\sum_{i=1}^n \lambda_i = 1, \lambda_i \geq0 \Biggr\}
\\
&&{} + \beta\inf\Biggl\{ \sum_{i=1}^n \lambda_i Z_i
\Bigm| x = \sum_{i=1}^n \lambda_i X_i,
\sum_{i=1}^n \lambda_i = 1, \lambda_i \geq0 \Biggr\}
\\
&=& \alpha g_{(X, Y)}(x) + \beta g_{(X, Z)}(x).
\end{eqnarray*}
As $\psi$ is nonincreasing and convex, we obtain
%
\begin{eqnarray}
\label{!convex}
&&
\int\psi\bigl( g_{(X,\alpha Y + \beta Z)}(x) \bigr) \,dx\nonumber\\
&&\qquad\leq
\int\psi\bigl( \alpha g_{(X, Y)}(x) + \beta g_{(X, Z)}(x) \bigr) \,dx
\nonumber\\[-8pt]\\[-8pt]
&&\qquad\leq\int\alpha\psi\bigl(g_{(X, Y)}(x)\bigr) + \beta\psi\bigl(g_{(X, Z)}(x)\bigr) \,dx
\nonumber\\
&&\qquad= \alpha\int\psi\bigl(g_{(X, Y)}(x)\bigr) \,dx
+ \beta\int\psi\bigl(g_{(X, Z)}(x)\bigr) \,dx\nonumber
\end{eqnarray}
as was required. Note that the integral is also finite whenever $Y$
has all components in the domain of $\psi$, due to the polyhedral
character of $g_{(X, Y)}(x)$ and the fact that $\psi$ is nonincreasing
and $\HH(X)$ is bounded. Otherwise, it may be equal only to $+\infty$;
hence $\Psi_{\mathcal{D}}$ is a proper convex function.
\begin{lemma}
\label{l:multi}
Suppose that $Y,Z$ are vectors in $\RR^d$ such that $Y=(y, \ldots, y)$
has constant components, and $Z$ is arbitrary. For any $\tau> 0$,
%
\begin{equation}
g_{(X,Y + \tau Z)}(x) = g_{(X,Y)}(x) + \tau g_{(X,Z)}(x)
= y + \tau g_{(X,Z)}(x).
\end{equation}
\end{lemma}
\begin{pf}
Note first that by the definition, $g_{(X,Y)}(x) = y$ identically on
$\HH(X)$ for constant $Y$; likewise, for every $x \in\HH(X)$,
\begin{eqnarray*}
g_{(X,Y + \tau Z)}(x) &=&
\inf\Biggl\{ \sum_{i=1}^n \lambda_i (Y_i + \tau Z_i)
\Bigm| x = \sum_{i=1}^n \lambda_i X_i,
\sum_{i=1}^n \lambda_i = 1, \lambda_i \geq0 \Biggr\}
\\
&=& \inf\Biggl\{ y + \tau\sum_{i=1}^n \lambda_i Z_i
\Bigm| x = \sum_{i=1}^n \lambda_i X_i,
\sum_{i=1}^n \lambda_i = 1, \lambda_i \geq0 \Biggr\}
\\
&=& y + \tau\inf\Biggl\{ \sum_{i=1}^n \lambda_i Z_i
\Bigm| x = \sum_{i=1}^n \lambda_i X_i,
\sum_{i=1}^n \lambda_i = 1, \lambda_i \geq0 \Biggr\}
\\
&=& g_{(X,Y)}(x) + \tau g_{(X, Z)}(x),
\end{eqnarray*}
proving the lemma.
\end{pf}

Choose a real number $y$ lying in the domain of $\psi$, and set
$Y=(y,\ldots,y)$. According to Lemma \ref{l:multi}, $g_{(X,Y)}(x)=y$
for every $x \in\HH(X)$; the function constant on $\HH(X)$ is convex,
hence $Y \in\mathcal{D}(X)$. Then
\[
\Phi_{\mathcal{D}}(Y) =
\frac{1}{n}\sum_{i=1}^n Y_i + \int\psi\bigl(g_{(X,Y)}(x)\bigr) \,dx
= y + \psi(y) \operatorname{Vol}(\HH(X)) < +\infty;
\]
therefore, $Y$ lies in the domain of $\Phi_{\mathcal{D}}$. For
arbitrary $Z \in\RR^n$, not identically $0$, and $\tau> 0$, we have
\begin{eqnarray*}
\Phi_{\mathcal{D}}(Y + \tau Z) &=&
\frac{1}{n}\sum_{i=1}^n (Y_i+\tau Z_i)
+ \int\psi\bigl(g_{(X,Y + \tau Z)}(x)\bigr) \,dx
\\
&=& y + \frac{\tau}{n}\sum_{i=1}^n Z_i
+ \int\psi\bigl(y + \tau g_{(X,Z)}(x)\bigr) \,dx
\\
&=& y + \tau\Biggl( \frac{1}{n}\sum_{i=1}^n Z_i
+ \int\frac{\psi(y + \tau g_{(X,Z)}(x))}{\tau} \,dx \Biggr).
\end{eqnarray*}
We know that $\Phi_{\mathcal{D}}$ is a convex function on a
finite-dimensional linear space $\RR^n$; to establish the existence of
its minimizer, it is sufficient to show that $\Phi_{\mathcal{D}}(Y
+ \tau Z) \to+\infty$ for $\tau\to+\infty$, which means that we
need to verify that the limit of the expression in the parentheses is
positive (possibly $+\infty$); see also \citet{HirLem93}, Remark
3.2.8. Let $E^-$, $E^0$, $E^+$ be sets in $\HH(X)$ where $g(x) =
g_{(X,Z)}(x)$ is, respectively, negative, zero or positive; we are to
examine the limit behavior of
%
\begin{eqnarray}
\label{!trint}
&&\frac{1}{n}\sum_{i=1}^n Z_i
+ \int_{E^-} \frac{\psi(y + \tau g(x))}{\tau} \,dx
\nonumber\\[-8pt]\\[-8pt]
&&\qquad{} + \int_{E^0} \frac{\psi(y)}{\tau} \,dx
+ \int_{E^+} \frac{\psi(y + \tau g(x))}{\tau} \,dx,\nonumber
\end{eqnarray}
when $\tau\to+\infty$. For the integral over $E^0$, the limit is
obviously zero. If $\psi$ satisfies assumptions (A1) and (A5), then
$\psi$ is nonincreasing and converging to $0$ when $\tau\to+\infty$;
for every $x \in E^+$ then $\psi(y+\tau g(x))/\tau$ monotonically
decreases with increasing $\tau$, hence the limit of the integral over
$E^+$ is zero as well. Finally, if $\psi$ satisfies also (A3), then
for every $x \in E^-$ the limit of $\psi(y+\tau g(x))/\tau$ is
$+\infty$; at the same time, the expression is bounded from below by
$0$, due to (A4). The application of the Fatou lemma then gives that
the limit of the integral over $E^-$ is $+\infty$, whenever $E^+$ has
positive Lebesgue measure.

The proof of the theorem is then finished by the examination of
possible alternatives. If the first term in (\ref{!trint}), the mean
of the $Z_i$'s, is positive, then the theorem is proved, as all other
terms in (\ref{!trint}) converge either to $0$ or $+\infty$. If the
first term of (\ref{!trint}) is negative or equal to zero, then there
must be some $Z_i < 0$ (the case with all $Z_i = 0$ is excluded). That
means that $g(x) = g_{(X,Z)}(x)$ is negative for some $x$; this
implies that $E^-$ has positive Lebesgue measure, and then the limit
of the second term in (\ref{!trint}), the integral over $E^-$, and
thus of the whole expression (\ref{!trint}) is $+\infty$. This proves
the theorem.

Under the strict convexity of $\psi$, the strict convexity of
$\Phi_{\mathcal{D}}$ follows (for appropriate $\alpha$, $\beta$) from
the second inequality in (\ref{!convex}), which becomes sharp---this
is due to the fact that the sharp inequality holds pointwise for all
$x$, and thus for the integrals as well. The strict convexity of
$\Phi_{\mathcal{D}}$ then implies the uniqueness of the solution.

Finally, functions $\psi$ satisfying (A1)--(A3), but not necessarily
(A5)
require some special considerations. For the integral over $E^-$, we
have to observe that for every $\tau> 0$ we have $\psi(y+\tau
g_{(X,Z)}(x)) \geq\psi$; if $\psi(y) < 0$, then
$\psi(y)/\tau\geq\psi(y)$ for every $\tau\geq1$, if $\psi(y)
\geq
0$ then $\psi(y)/\tau\geq0$ for every $\tau> 0$. Hence we have also
in this case an integrable constant minorant [due to the fact that
$\HH(X)$ has finite Lebesgue measure]; this justifies the limit
transition via the Fatou lemma. Finally, for the integral over $E^+$,
we need to assume that the limit of the integrand is $0$ for every
$x$, and find an integrable minorant; this may be related to the
existence of the integral of the second term in (\ref{!dumb}).
\end{pf*}
\begin{pf*}{Proof of Theorem \ref{t:fish}}
The proof relies on the application of what is called Fenchel's
inequality by Rockafellar [(\citeyear{Roc70}), page 105] or the (generalized) Young
inequality by Hardy, Littlewood and P\'olya [(\citeyear{HLP}), Section 4.8],
or Zeidler [(\citeyear{Zei85}), Section~51.1]. The
inequality says says that for arbitrary $x,y$ and convex function~$\psi$
\[
xy \leq\psi(x) + \psi^{\ast}(y).
\]
Applied pointwise to $x = -f_0$ and $y=\chi^{-1}(-f)$, the inequality yields
\[
(-f_0) \chi^{-1}(-f) \leq\psi^{\ast}(-f_0) + \psi(\chi^{-1}(-f)),
\]
which is equivalent to the nonnegativity of the integrand
in (\ref{!pophub}). For $f=f_0$, the equality (\ref{!legenf0}) implies
that
\[
f_0 \chi^{-1}(-f_0) + \psi^{\ast}(-f_0) + \psi(\chi^{-1}(-f_0))
= - \psi^{\ast}(-f_0) + \psi^{\ast}(-f_0) = 0,
\]
which proves the theorem.

For functions not satisfying (A5), integrability of $f_0 \chi^{-1}
(-f_0)$ is
no longer equivalent to that of $-\psi^{\ast}(-f_0)$. However, if we
assume the integrability of the latter, then the proof can be carried through
in the same way.
\end{pf*}

\vspace*{-12pt}

\section{Computational details}\label{appB}

Our computational objective is to provide a unified algorithmic
strategy for solving the entire class of problems described above.
Interior point methods designed for general convex programming and
capable of exploiting the inherently sparse structure of the resulting
linear algebra offer a powerful, general approach. We have employed
two such implementations throughout our development process: the PDCO
algorithm of \citet{saunders04}, and the MOSEK implementation of
\citet{And06}.

Our generic primal problem (\ref{!dumb}) involves minimizing an
objective function consisting of a linear component, representing
likelihood or some generalized notion of fidelity, plus a nonlinear
component, representing the integrability constraint. Minimization
is then subject to a cone constraint imposing convexity. We will
first describe our procedure for enforcing convexity, and then turn
to the integrability constraint.

\subsection{\texorpdfstring{The convexity constraint.}{The convexity constraint}}

In dimension one convexity of piecewise linear functions can be
imposed easily by enforcing linear inequality constraints on a set of
function values, $\gamma_i = g(\xi_i)$ at selected points
$\xi_1,\xi_2,\ldots,\xi_m$. For ordered $\xi_i$'s, the convex cone
constraint can be written as $D \gamma\geq0$ for a tridiagonal
matrix $D$ that does second differencing, adapted to the possible
unequal spacing of the $\xi_i$'s.

In dimension two, enforcing convexity becomes more of a challenge.
Ideally, we would utilize knowledge of the polyhedral character of the
optimal $g$, established by Theorem \ref{t:first}, and implying that
the optimal $g$ is piecewise linear over \textit{some} triangulation of
the observations $X_i \in\RR^2$. Once we knew the triangulation, it
is again straightforward to impose convexity: each interior edge of
the triangulation generates one linear inequality on the coefficients
$\gamma$. Unfortunately, the complexity of traveling over a binary
tree of possible triangulations of the observed points makes finding
the optimal one difficult. The algorithm implemented in the
\texttt{logConcDEAD} package of \citet{CGS}, for computing log-concave
estimates, exploits special features of the log-concave MLE problem
and thus does not appear to be easily generalizable to our other
settings. Finite-element methods involving fixed (Delaunay)
triangulation of an expanded set of vertices were also ultimately
deemed unsatisfactory.

A superior choice, one that circumvents the difficulties of the
finite-element, fixed triangulation approach, relies on finite
differences. Convexity is imposed directly at points on a regular
rectangular grid using finite-differences to compute the discrete
Hessian:
\begin{eqnarray*}
H_{11} (\xi_1,\xi_2)
&=& g(\xi_1+\delta,\xi_2) -2 g(\xi_1,\xi_2) + g(\xi_1-\delta,\xi
_2), \\
H_{22} (\xi_1,\xi_2)
&=& g(\xi_1,\xi_2+\delta) -2 g(\xi_1,\xi_2) + g(\xi_1,\xi
_2-\delta), \\
H_{12} (\xi_1,\xi_2)
&=& [g(\xi_1+\delta,\xi_2+\delta) - g(\xi_1+\delta,\xi_2-\delta
)\\
&&\hspace*{1pt}{} - g(\xi_1-\delta,\xi_2+\delta) + g(\xi_1-\delta,\xi
_2-\delta)]/4, \\
H_{21} (\xi_1,\xi_2) &=& H_{12} (\xi_1,\xi_2).
\end{eqnarray*}
Convexity is then enforced by imposing positive semidefiniteness.
These con\-straints---convexity at each of the grid points
$(\xi_{1i},\xi_{2i})$---produce a semi-definite programming problem.
In the bivariate setting the semi-definiteness of each $H$ can be
reformulated as a rotated quadratic cone constraint; we need only
constrain the signs of the diagonal elements of $H$ and its
determinant. This simplifies the implementation of the Hellinger
estimator in MOSEK. For the relatively fine grid used for Figure \ref{fig7}
solution requires about 25 seconds, considerably quicker than the
log-concave estimate of Figure \ref{fig8} computed with the implementation of
\citet{CGS}.

\subsection{\texorpdfstring{The integrability constraint.}{The integrability constraint}}

For certain special $\psi$, one can evaluate the integral term $\int
\psi(g(x)) \,dx$ in the objective function of (\ref{!dumb})
explicitly---as was done for $\psi(g) = e^{-g}$ by \citet{CSS}. While
such a strategy may also be possible for certain other specific
$\psi$, we adopt a more pragmatic approach based on a straightforward
Riemannian approximation
%
\begin{equation}
\label{!riemann}
\int_{\HH(X)} \psi(g(x))\,dx \approx
\sum_{i=1}^m \psi(g (\xi_i)) s_i.
\end{equation}
Here, $s_i$ are weights derived from the configuration of $\xi_i$. Of
course, with only a modest number of $\xi_i$'s such an approximation
may be poor; in dimension one we therefore augment the initial
collection of the observed data $X_1, \ldots, X_n$ by filling the
gaps between their order statistics by further grid points, to ensure
that the resulting grid (not necessarily uniformly spaced) and
consisting of the observed data points as well as the new grid points,
provides a sufficiently accurate approximation (\ref{!riemann}). The
$s_i$'s are then simply the averages of the adjacent spacings between
the ordered $\xi_i$'s. Given the size of problems modern optimization
software can successfully handle, it is no problem to add an abundance
of new points in dimension one.

In dimension two, the approximation (\ref{!riemann}) is based on the
uniformly spaced grid of the points used in the finite-difference
approach described in the previous subsection. As the original data
points $X_i$ may no longer lie among the grid points~$\xi_i$, we have
to modify the fidelity component of the objective function: instead of
obtaining $g(X_i)$ directly, we obtain it via linear interpolation
from the values of $g$ at the vertices of the rectangles enclosing
$X_i$. As long as the grid is sufficiently fine, the difference is
minimal. We use this approach often also in dimension one, as it
provides better numerical stability especially for fine grids and
large data sets.

\subsection{\texorpdfstring{Discrete duality.}{Discrete duality}}

Adopting the procedures described above, we can write the
finite-dimensional version of the primal problem as
\renewcommand{\theequation}{$\mathrm{P}$}
\begin{equation}\label{equP}
\{ w^\top L \gamma+ s^\top\Psi(\gamma) | D \gamma\geq0 \} = \min!,
\end{equation}
where $\Psi(\gamma)$ denotes now the $m$-vector with typical element
$\psi(g(\xi_i)) = \psi(\gamma_i)$, $L$ is an ``evaluation operator''
which either selects the data elements from $\gamma$, or performs the
appropriate linear interpolation from the neighboring ones, so that $L
\gamma$ denotes the $n$-vector with typical element, $g(X_i)$, and $w$
is an $n$-vector of observation weights, typically $w_i \equiv1/n$.

Associated with the primal problem (\ref{equP}) is the dual problem
\renewcommand{\theequation}{$\mathrm{D}$}
\begin{equation}\label{equD}
\{ -s^\top\Psi^{\ast} (- \phi) | S \phi= -w^\top L + D^\top\eta,
\phi\geq0, D^\top\eta\geq0 \} = \max!.
\end{equation}
Here, $\eta$ is an $m$-vector of dual variables and $\phi$ is an
$m$-vector of function values representing the density evaluated at
the $\xi_i$'s, and $S = \diag(s)$. The vector $\Psi^{\ast}$ is the
convex conjugate of $\Psi$ defined coordinate-wise with typical
element $\psi^{\ast} (y) = \sup_x \{ y x - \psi(x) \}$. Problems (\ref{equP})
and (\ref{equD}) are strongly dual in the sense of the following result, which
may viewed as the discrete counterpart of Theorem~\ref{t:dual}.
\begin{proposition}
If $\psi$ is convex and differentiable on the interior $\II$ of its
domain, then the corresponding solutions of (\ref{equP}) and (\ref{equD}) satisfy
\renewcommand{\theequation}{$\mathrm{E}$}
\begin{equation}\label{equE}
f(\xi_i) = \psi' (g (\xi_i)) \qquad\mbox{for } i = 1, \ldots, m,
\end{equation}
whenever the elements of g are from $\II$ and the elements of $f$ are
from the image of $\II$ under $\psi'$.
\end{proposition}

For $\Psi(x)$ with typical element $\psi(x) = e^{-x}$ we have
$\Psi^{\ast}$ with elements $\psi^{\ast} (y) = -y \log y + y$, so the
dual problem corresponding to maximum likelihood can be interpreted as
maximizing the Shannon entropy of the estimated density subject to the
constraints appearing in (\ref{equD}). Since $g$ was interpreted in (\ref{equP}) as
$\log f$, this result justifies our interpretation of solutions of (\ref{equD})
as densities provided that they satisfy our integrability condition.
This is easily verified and thus justifies the implicit Lagrange
multiplier of one on the integrability constraint in (\ref{equP}), giving a
discrete counterpart of Theorem \ref{t:dual}.
\begin{proposition}
Let $\iota$ denote an $m$-vector of ones, and suppose in (\ref{equP}) that
$w^\top L \iota=1$ and $D \iota= 0$. Then solutions $\phi$ of (\ref{equD})
satisfy $s^\top\phi= 1$ and $\phi\geq0$.
\end{proposition}

The crucial element of the proof is that the differencing operator D
annihilates the constant vector and therefore the result extends
immediately to other norm-type penalties as well as to the other
entropy objectives that we have discussed. Indeed, since the second
difference operator representing our convexity constraint annihilates
any affine function it follows by the same argument that the mean of
the estimated density also coincides with the sample mean of the
observed $X_i$'s.
\end{appendix}

\section*{\texorpdfstring{Acknowledgments.}{Acknowledgments}}

We are grateful to Lutz D\"umbgen, Kaspar Rufibach, Guenther Walther
and Jon Wellner for sending us preprints of their work, to the
referees for their very constructive comments, and to Mu Lin for help
with the bright star example and Fisher consistency proof.

\printaddresses

\end{document}